\documentclass{article}\usepackage{jkas2}
\runningauthor{CLARKE}
\runningtitle{Faraday Observations of Cluster Fields}
\beginpage{1}
\endpage{5}

\begin{document}

\font\twelvei = cmmi10 scaled\magstep1 
       \font\teni = cmmi10 \font\seveni = cmmi7
\font\mbf = cmmib10 scaled\magstep1
       \font\mbfs = cmmib10 \font\mbfss = cmmib10 scaled 833
\font\msybf = cmbsy10 scaled\magstep1
       \font\msybfs = cmbsy10 \font\msybfss = cmbsy10 scaled 833
\textfont1 = \twelvei
       \scriptfont1 = \twelvei \scriptscriptfont1 = \teni
       \def\mit{\fam1 }
\textfont9 = \mbf
       \scriptfont9 = \mbfs \scriptscriptfont9 = \mbfss
       \def\bmit{\fam9 }
\textfont10 = \msybf
       \scriptfont10 = \msybfs \scriptscriptfont10 = \msybfss
       \def\bmsy{\fam10 }

\def\etal{{\it et al.~}}
\def\eg{{\it e.g.,~}}
\def\ie{{\it i.e.,~}}
\def\lsim{\raise0.3ex\hbox{$<$}\kern-0.75em{\lower0.65ex\hbox{$\sim$}}}
\def\gsim{\raise0.3ex\hbox{$>$}\kern-0.75em{\lower0.65ex\hbox{$\sim$}}}

\title{Faraday Rotation Observations of Magnetic Fields in Galaxy Clusters}

\author{Tracy E.\ Clarke}
\address{Department of Astronomy,
University of Virginia, P.\ O.\ Box 3818, Charlottesville, VA  22903-0818  USA\\
{\it E-mail: tclarke@virginia.edu}}

%


\address{\normalsize{\it (Received October 31, 2004; Accepted December 1,2004)}}

\abstract{ The presence of magnetic fields in the intracluster medium
in clusters of galaxies has been revealed through several different
observational techniques. These fields may be dynamically important in
clusters as they will provide additional pressure support to the
intracluster medium as well as inhibit transport mechanisms such as
thermal conduction. Here, we review the current observational state of
Faraday rotation measure studies of the cluster fields. The fields are
generally found to be a few to 10 $\mu$G in non-cooling core clusters
and ordered on scales of $10-20$ kpc. Studies of sources at large
impact parameters show that the magnetic fields extend from cluster
cores to radii of at least 500 kpc. In central regions of cooling core
systems the field strengths are often somewhat higher ($10-40\ \mu$G)
and appear to be ordered on smaller scales of a few to 10 kpc. We also
review some of the recent work on interpreting Faraday rotation
measure observations through theory and numerical simulations. These
techniques allow us to build up a much more detailed view of the
strength and topology of the fields. }


\maketitle

\section {Introduction}

Magnetic fields are common to nearly all astrophysical phenomena, thus
it is not surprising to find that they are an important constituent of
galaxy clusters, and therefore a major topic for this
meeting. Intracluster magnetic fields provide an additional form of
pressure support to the cluster gas, as well as inhibiting transport
processes such as cosmic ray propagation and thermal conduction in the
cluster. The presence of intracluster magnetic fields has been
established through a number of observational approaches. This paper
concentrates on Faraday rotation measure (RM) observations of the
magnetic fields, mainly on the observational status of the RM studies,
although we also briefly summarize some of the important work on
modeling the fields in the intracluster medium (ICM) based on RM
observations. Other techniques to study the intracluster magnetic
fields are presented elsewhere in these proceedings. For detailed
recent reviews of intracluster magnetic fields, the reader is referred
to Govoni \& Feretti (2004), Clarke (2003), Carilli \& Taylor (2002),
Widrow (2002), and Kronberg (1994).

\section{Faraday Rotation Measure Overview}

One of the most direct measurements of intracluster magnetic fields is
through the effect of the fields on the propagation of linearly
polarized radiation. As the polarized emission from a radio source
passes through a magnetized, ionized plasma, the plane of polarization
will be rotated due to the different phase velocities of the two
opposite-hand polarization modes. This Faraday rotation effect will
rotate the intrinsic position angle of the emission by an addition
term given by the rotation measure (RM):
\begin{equation}
RM= 811.9\int_0^L n_{\rm e} B_\| d\ell \ \ \ {\rm rad/m^2}
\label{eq:rm}
\end{equation}
where $n_{\rm e}$ is the electron density in ${\rm cm^{-3}}$, $B$ is
the line of sight magnetic field strength in $\mu$G, and $L$ is the
path length through the Faraday rotating medium in kpc. The change in
position angle of the radiation can be written as:
\begin{equation}
\Delta\chi = \chi - \chi_0 = RM\,\lambda^2 \ \ {\rm radians}
\end{equation}
where $\chi$ is the observed position angle at wavelength $\lambda$
and $\chi_0$ is the intrinsic position angle of the polarized
emission. Faraday rotation measure studies require observations at
three or more wavelengths in order to remove the $n\pi$ ambiguity
resulting from the fact that measured position angle is only a
pseudo-vector lying in the range of $0 < \chi < \pi$ radians. The
measurements of this angle are uncertain by $\pm n\pi$ due to the
unknown number of half rotations of the polarization angle between the
source and observer. By carefully selecting a set of observational
wavelengths, it is possible to remove the ambiguities and determine
accurate rotation measures. A more detailed discussion of the $n\pi$
ambiguities is given by Ruzmaikin \& Sokoloff (1979).

The observed rotation measure is a linear sum of all contributions
along the line of sight between the radio source and the
observer. Contributions to the observed RM are mainly limited to the
region local to the radio source, the intracluster medium, and the
Milky Way galaxy. Since it is the cluster contribution which we are
interested in, it is important to understand (and limit) the
contributions from the other sources of Faraday rotation. The Galactic
rotation measure has been studied by many authors, including a recent
study by Frick et al.\ (2001). In general, these studies find that the
Galactic contribution to the RM can be up to a few hundred radians
m$^{-2}$ at low galactic latitudes, but this contribution quickly
drops to the level of a few radians m$^{-2}$ above a Galactic latitude
of $\sim$ 20$^\circ$. The contribution local to the source appears to
be generally small based on observations such as the Laing-Garrington
effect where the more distant radio lobe is more depolarized than the
lobe pointing toward us (Laing 1988, Garrington et al.\ 1988, 1991),
as well as the radial trend seen in statistical Faraday studies (see
\S~\ref{sect:stat}), and the results of gradient alignment statistics
(En{\ss}lin et al.\ 2003). For observations limited to high Galactic
latitudes, the majority of the observed Faraday rotation for sources
viewed through galaxy clusters will be contributed by the intracluster
magnetic fields.

An estimate of the magnetic field strength from Equation~\ref{eq:rm}
requires some assumptions to be made about the geometry of the
intracluster fields. The simplest case would be a uniform slab
geometry where both the direction and strength of the magnetic field
are constant through the cluster medium. This geometry provides a
lower limit on the magnetic field strength. As discussed in
\S~\ref{sect:map}, Faraday maps of extended sources often show
fluctuations on scales much smaller than the radio source extent. An
alternative approach to the field topology is to assume that the
cluster is composed of cells of uniform strength but random field
orientation along the line of sight. Generally the magnetic field
scale length for this tangled cell model is taken to be the observed
RM fluctuation scale. Note, however, that recent work by En{\ss}lin \&
Vogt (2003) and Vogt \& En{\ss}lin (2003) has shown that the RM
fluctuation scale overestimates the true field coherence length, and
hence underestimates the field strength in this simple model.
 
\section{Extended and Polarized Radio Sources}
\label{sect:map}

\subsection{Faraday Mapping}

Details on the topology of the Faraday rotating medium require high
resolution Faraday mapping of extended, polarized radio sources. The
majority of these studies have been undertaken for extended sources
embedded in the cores of cooling flow clusters (Ge \& Owen 1993, 1994;
Taylor \& Perley 1993; Taylor, Fabian, \& Allen 2002). 

One of the first objects targeted for high resolution Faraday studies
was Cygnus A located in the core of a dense cluster. Dreher, Carilli
\& Perley (1987) found large rotation measure gradients of up to 600
radians m$^{-2}$ across the lobes of the radio galaxy, conclusively
ruling out the possibility of a Galactic origin to the RMs. Perhaps
more important was the fact that they found the position angle varied
by more than $\pi$ radians and was well fit to a $\lambda^2$ relation
over a very large range of frequencies. Combining that with the fact
that the polarization fraction remained constant over the region
allowed them to rule out the possibility that the RMs were due to
thermal gas mixed into the radio emitting plasma. These were the first
observations to conclusively show that the observed RMs must be coming
mainly from an external Faraday screen which could not be
Galactic. After an analysis of a number of possibilities, Dreher et
al.\ (1987) concluded that the most likely origin for the RMs was the
intracluster medium. Based on simple assumptions of field topology,
they estimate the intracluster magnetic field strength to be $2-10\ 
\mu$G for this cluster. 

\begin{figure}[t]
\vskip 0.11truein
\centerline{\epsfysize=\columnwidth\epsfbox{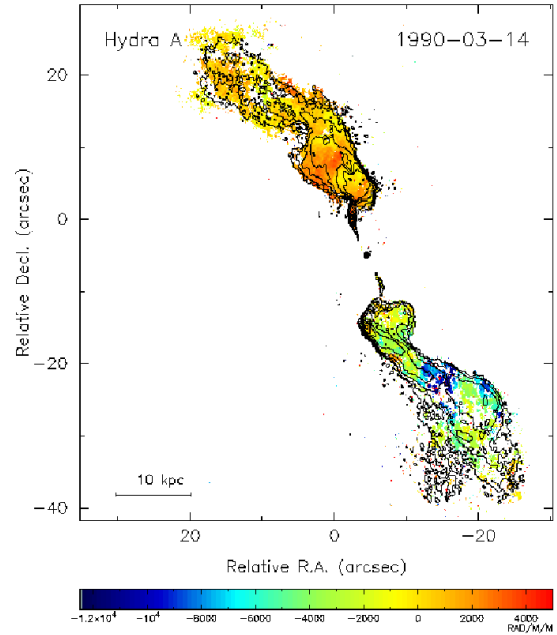}}
\vskip -0.1truein
\caption{ Faraday rotation measure map of Hydra A in color with total
intensity contours overlaid (Taylor \& Perley 1993). The northern lobe
is mainly positive (field toward observer) while the southern lobe is
mainly negative (field away from observer) indicating organized
structures on scales of 50 kpc. The map also shows small scale
fluctuations on scales of $5-10$ kpc, indicating the presence of a
tangled field component.  }
\label{fig:hydra}
\end{figure}

Observations of Hydra A by Taylor \& Perley (1993) show mainly
positive RMs in the northern lobe and negative values across the
southern lobe (Figure~\ref{fig:hydra}). This RM structure requires a
field geometry that is ordered on scales of 50 kpc in the cluster. The
Faraday map of Hydra A also shows small scale fluctuations (scale of
$5-10$ kpc) superimposed on the lobes. Based on these observations
Taylor \& Perley (1993) estimate the uniform component of the field to
be $\sim$ 7 $\mu$G and the tangled component to be $\sim$ 40 $\mu$G.

\begin{figure}[t]
\vskip 0truein
\centerline{\epsfysize=\columnwidth\epsfbox{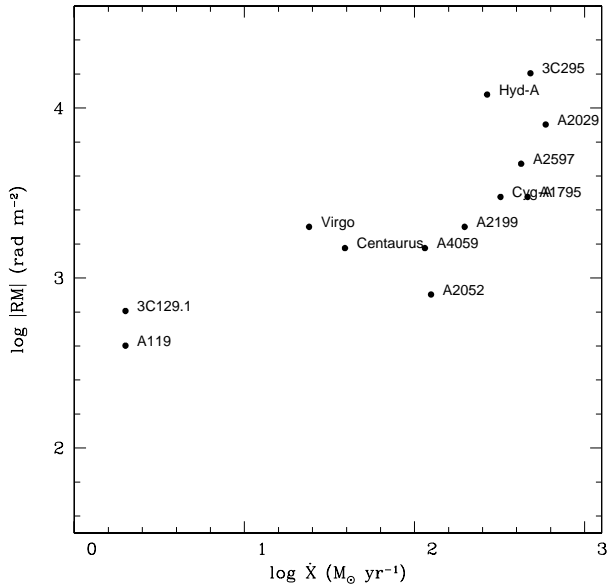}}
\vskip -0.1truein
\caption{ Plot of maximum absolute rotation measure as a function of
the cluster cooling rate from Taylor, Fabian, \& Allen (2002). They
point out that all known high rotation measure sources at z $<$ 0.4
are included in this sample. These data are also consistent with all
radio sources embedded in cooling core clusters having extreme RMs. }
\label{fig:RM_X}
\end{figure}

More generally, the Faraday studies of cooling core sources find very
high rotation measures ($|RM| > 800$ rad/m$^2$), with the magnitude of
the RMs appearing roughly proportional to the cooling flow rate ($\dot
X$) (Taylor et al.\ 2002). This correlation (shown in
Figure~\ref{fig:RM_X}) is expected as both the RM and $\dot X$ depend
on the electron density to positive powers. The RM fluctuations seen
in maps of cooling core clusters are on scales of a few to 10 kpc,
resulting in a tangled cell estimate of $10-40\ \mu$G (Carilli \&
Taylor 2002).

Extended, polarized radio sources have also been studied in
non-cooling core clusters. Eilek \& Owen (2002) studied the RM
properties of 3C 75 in the core of Abell 400, and 3C 465 in the core
of Abell 2634. Both radio sources extend $>$ 100 kpc in the cluster
cores and were observed with resolutions $\sim$ 2 kpc. The Faraday
maps show patches of RM fluctuations on scales of order $10-20$
kpc. If this scale is taken to be the tangled magnetic field reversal
scale, Eilek \& Owen estimate that the cluster core magnetic field
strength is in the range of $1-4$ $\mu$G. 

A more detailed picture of magnetic fields in a single cluster can be
obtained by mapping several polarized radio sources viewed at
different impact parameters to the cluster core. In Abell 119, Feretti
et al.\ (1999) studied the rotation measures of 3 extended polarized
sources located within the cluster at impact parameters of 115, 300,
1010 kpc. Overall they found that the magnitude of the maximum RM, the
dispersion in the RM, and the depolarization at long wavelengths all
decrease with increasing distance from the cluster center. This is
consistent with the idea that the Faraday screen in front of the
sources is tangled on small scales ($<$ 4 kpc) in the central regions
of the cluster. From a tangled cell analysis they estimate the
magnetic field strength to be in the range of $12-17\ \mu$G, although
they note that such a field strength is unlikely for the source at the
largest impact parameter as the magnetic pressure would exceed the
thermal pressure in this region. Indeed, a re-analysis of the Abell
119 data by Dolag et al.\ (2001) finds that the fields scale as
$n_{\rm e}^{0.9}$, significantly steeper than the $n_{\rm e}^{0.5}$
expected for constant scaling between the thermal and magnetic
pressures in the ICM.

\subsection{Theory and Simulations} 

Recently, significant effort has been put into extending the analysis
of Faraday maps beyond the uniform slab and tangled cell models
described above. Detailed cosmological simulations including magnetic
fields are discussed elsewhere in these proceedings by Dolag. Here, we
concentrate on a brief description of methods that are being applied
to observational Faraday maps. 

En{\ss}lin \& Vogt (2003) describe the details of a real space and
Fourier space analysis of high resolution Faraday rotation measure
maps. Using the techniques they describe, they are able to to
determine the strength, energy spectrum, and magnetic field
autocorrelation length ($\lambda_B$) from observational data. They
find that $\lambda_B$ can differ from the (observed) RM fluctuation
scale that is commonly equated to the magnetic field scale length. In
a typical astrophysical plasma they find that the RM scale length is
generally larger than $\lambda_B$, thus the field strengths may be
underestimated in the simple tangled cell model. Vogt \& En{\ss}lin
(2003) applied their techniques to Faraday maps of 3C 75, 3C 465, and
Hydra A, and found that the RM fluctuation scale is generally $2-4$
times larger than $\lambda_B$ for these sources. They determine
magnetic to thermal pressures between 1\% and 20\% in the cores of
these clusters, and conclude that the magnetic fields may have an
impact on the cluster dynamics. The reader is referred to the
contribution by Vogt \& En{\ss}lin in these proceedings for further
details.

Improvements in the modeling of the intracluster magnetic fields have
also been made recently. Work by Murgia et al.\ (2004) has applied
Monte Carlo methods to simulate random 3D magnetic fields in
clusters. They select systems which contain details of the
polarization structure of several radio galaxies viewed through the
cluster and, in some cases, clusters which also have information on
the total intensity distribution of diffuse radio halos. They study
the different power spectrum spectral indices to determine what value
provides the best fit to the polarization characteristics of the
sources as well as the surface brightness distribution and
polarization limits of the diffuse radio halo emission if
applicable. Based on their analysis of Abell 119, they find mean
intracluster magnetic field strengths of 2 $\mu$G are required to
reproduce the observed RM structure of the embedded sources. Murgia et
al.\ (2004) also find that the data require a relatively flat power
spectrum of $n=2$, indicating that much of the magnetic field energy
density is on small scales in the intracluster medium. They also find
that the field analysis using the power spectrum results in a field
strength that is a factor of roughly 2 lower than that obtained from
the standard RM fluctuation scale analysis.
 
\section{Statistical Faraday Studies}
\label{sect:stat}

Ideally, the radial distribution of the strength and topology of
intracluster magnetic fields would be studied on an individual cluster
basis with a large number of Faraday probes at different cluster
impact parameters. Unfortunately, the analysis of a large number of
clusters with many RM probes per cluster is currently unfeasible due
to the sensitivity limits of available radio telescopes. An
alternative approach to studying the distribution of intracluster
magnetic fields is to obtain the RM for a number of probes viewed
through a carefully selected sample of galaxy clusters. By choosing a
sample of clusters with similar characteristics (e.g.\ the lack of a
central cooling core), a statistical approach can be taken to study
the magnetic fields. In addition to the Faraday observations, the
statistical analysis requires details of the distribution of the
thermal gas in each cluster. Until recently, this type of analysis was
difficult due to a paucity of X-ray data required to obtain individual
electron density measurements for the target clusters. Statistical
studies such as those by Lawler \& Dennison (1982) or Kim, Tribble, \&
Kronberg (1991) used average properties over a sample of clusters to
determine the electron density distribution as many of the systems
under study did not have X-ray observations available. This approach
is not ideal since the clusters covered a very large range of richness
and morphology, thus the true electron density along a line of sight
could be significantly different from the assumed universal
profile. The era of sensitive, long-lived X-ray satellites such as
$ROSAT$, $Chandra$ and $XMM$ is now providing the necessary details of
individual cluster electron density distributions for determining the
magnetic field strength along individual sight-lines through a large
number of galaxy clusters.

The first large statistical Faraday study to obtain individual
electron density profiles toward each target cluster was undertaken by
Clarke, Kronberg, \& B\"ohringer (2001). Their study contained 16 low
redshift clusters which were selected to have bright (${\rm L_x >
5\times 10^{42}\ ergs\ s^{-1}}$), extended X-ray emission in the ROSAT
$0.1-2.4$ keV band. The cluster sample was further limited to high
Galactic latitudes ($|b|\ge 20^\circ$) to avoid contamination by the
Galactic magnetic field, low redshift ($z\le$ 0.1) to provide a large
angular extent on the sky for obtaining RM probes, and each cluster was
required to contain at least one linearly polarized radio source
viewed through the X-ray emitting ICM. In addition to these {\it
cluster} probes viewed through the ICM, a second set of polarized {\it
control} sources was observed toward each cluster at impact parameters
beyond the detectable X-ray emission in order to help determine the
Galactic contribution in the direction of each cluster. Target radio
sources were further constrained to provide sight-lines probing a
large range of cluster impact parameters in order to investigate the
radial extent of intracluster magnetic fields. Radio sources in the
{\it cluster} and {\it control} samples were observed with the NRAO
VLA at four to six wavelengths each within the 20 and 6 cm bands. The
observing wavelengths were selected to provide unambiguous RMs within
the range $|RM| \le 2600\ {\rm rad\ m^{-2}}$. X-ray observations of
each cluster were retrieved from the ROSAT Data Archive. Thirteen of the
clusters were observed in Pointed Observation mode, while the
remaining three clusters were extracted from the ROSAT All-Sky Survey
(RASS) archive. Details of the cluster sample including radio and
X-ray reductions are presented in Clarke et al.\ (2001).

\begin{figure}[t]
\vskip 0truein
\centerline{\epsfysize=\columnwidth\epsfbox{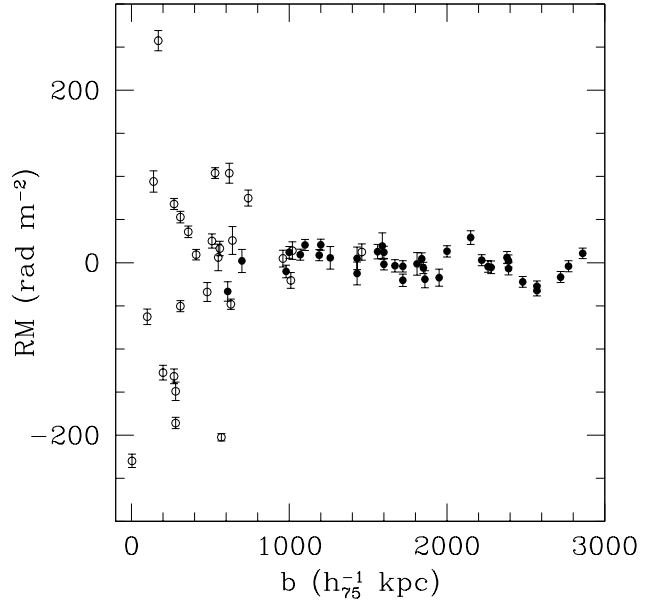}}
\vskip -0.1truein
\caption{ Rotation measure (corrected for the Galactic contribution)
plotted as a function of source impact parameter in kiloparsecs for 16
clusters from Clarke et al.\ (2001). Open points are the {\it cluster}
probes while filled points show the {\it control} sample. Note the
clear increase in the width of the RM distribution toward smaller
impact parameter, indicating the presence of intracluster magnetic
fields.  }
\label{fig:RRM_thesis}
\end{figure}

The distribution of Faraday rotation measures (corrected for the
Galactic rotation measure contribution) is shown in
Figure~\ref{fig:RRM_thesis}. This figure clearly shows a broadening of
the RM distribution toward small impact parameter, consistent with
excess Faraday rotation in the cluster sample out to impact parameters
of more than 500 kpc. Using a simple uniform slab model for the
cluster field provided lower limits on the field estimates of
$0.5-3.0\ \mu$G. Analysis of several extended radio sources in the
sample revealed RM fluctuation scales of order $10-20$ kpc, resulting
in tangled cell field estimates of $5-10\ \mu$G. 

Statistical studies also allow us to address the question of the
location of the Faraday rotating medium. Although the majority of the
observational evidence points toward fields located in the
intracluster medium (Carilli \& Taylor 2002), a contribution from a
region local to embedded radio sources cannot be completely ruled out
(Bicknell, Cameron, \& Gingold 1990, Rudnick \& Blundell 2003, Rudnick
these proceedings; but also see Carilli \& Taylor 2002, and En{\ss}lin
et al.\ 2003). Clarke et al.\ (in preparation) have expanded the
statistical Faraday sample to include high redshift background radio
probes viewed through low redshift clusters. Although redshifts are
not available for all targets, it is possible to tag individual
sources as possible clusters members or background sources based on
the Second Generation Palomar Survey images. For each candidate which
did not have a redshift, a source was only labeled background if all
candidate optical counterparts would have $M_R > -21$ if they were at
the redshift of the cluster (F.\ Owen, private communication). In
fact, all unidentified sources would have $M_R > -19.5$ if they were
cluster members, thus their identification as background sources is
expected to be reliable. This approach to separating embedded and
background probes is conservative, and will likely overestimate the
number of embedded sources. 
The F-test rejects the null hypothesis of similar variances for the
embedded (or background) and control samples, thus indicating that
both the embedded and background samples are seeing excess
intracluster Faraday rotation.

\begin{figure}[t]
\vskip 0truein
\centerline{\epsfysize=\columnwidth\epsfbox{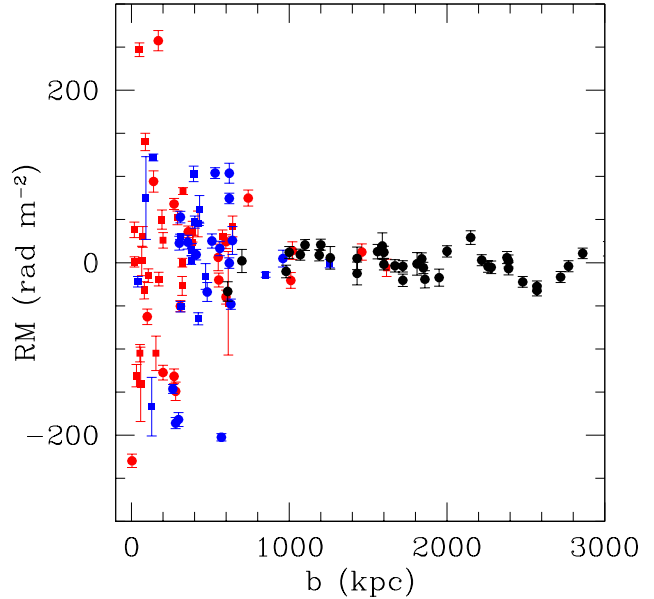}}
\vskip -0.1truein
\caption{Expanded RM sample from Clarke et al.\ (in preparation)
including data from the literature, separated into embedded (red),
background (blue), and control (black) samples. The increase in the
width of the RM distribution toward smaller impact parameter is
clearly visible for both the embedded and the background samples. }
\label{fig:RRM_comb}
\end{figure}

Using the identification techniques described above, additional
Faraday probes from the sample of Kim, Tribble, \& Kronberg (1991) can
be added to the Clarke et al.\ (2001) sample to obtain a combined data
set shown in Figure~\ref{fig:RRM_comb}. Applying the
Kolmogorov-Smirnov test on the combined samples rejects the null
hypothesis of the sources being drawn from the control sample at $>$
99.7\% for the background sample and $>$ 99.9\% for the embedded
sample. In addition, the Spearman Rank test shows that the background
RM sample is anti-correlated with impact parameter at high
probability, again indicating that the excess is due to the
intracluster magnetic field. The split RM samples provide strong
evidence that both the embedded and background radio probes are
sampling intracluster magnetic fields, and not a local Faraday screen
in the vicinity of the radio source.

\section{Summary}

Faraday rotation measure probes of intracluster magnetic fields show
evidence of cluster fields in both cooling and non-cooling core
clusters. In addition, several studies show that the fields extend to distances of
at least 500 kpc from the cluster cores. Magnetic field strengths in
non-cooling core clusters are generally a few $\mu$G and RM maps show
fluctuations on scales of $10-20$ kpc. Estimates of field strengths in
cooling core clusters are generally about an order of magnitude
larger, with the RM fluctuation scale being roughly an order of
magnitude smaller. The separation of Faraday probes into embedded and
background samples shows similar RM distributions for both samples,
indicating that the majority of the RM excess is from fields in the
intracluster medium. More detailed approaches are being explored to
translate the observational data into estimates of the field strength
and topology. Depending on the techniques used, these studies suggest
that the field scales and estimates from previous work may be off be a
factor of 2. 

The current instrumentation only allows us to measure the polarimetry
of a few background probes viewed through individual clusters due to
sensitivity limitations. In the near future, the EVLA will greatly
expand our capabilities and allow us to reach dozens of background
sources per cluster for many systems. This will permit the study of
the radial distribution of the RM properties across single clusters
and the comparison of these distributions for clusters in different
evolutionary stages to allow us to search for correlations between the
magnetic field and cluster properties. Deeper searches will also be
possible to detect (or place lower limits on) the polarization
associated with diffuse radio halos, and the improved sensitivity and
resolution will also allow more detailed studies of extended radio
galaxies within and behind clusters

\acknowledgements{ I would like to thank my collaborators Hans
B\"ohringer, Torsten En{\ss}lin, Phil Kronberg, Craig Sarazin, and
Corina Vogt for their efforts in this work. I am grateful to Luigina
Feretti, Federica Govoni, and Greg Taylor for interesting discussions
on this field. I would also like to thank the conference organizers
for enabling such a great scientific and cultural experience. The
National Radio Astronomy Observatory is a facility of the National
Science Foundation operated under a cooperative agreement by
Associated Universities, Inc. The ROSAT Data Archive is maintained by
the Max-Planck-Institut f\"ur extraterrestrische Physik (MPE) at
Garching, Germany. }

\end{document}